\documentclass[12pt]{article}
\usepackage{graphicx}
\usepackage{amsmath}
\usepackage{bm}% bold math
\def\Vec#1{\mbox{\boldmath $#1$}}

\def\itmb{\begin{itemize}}
\def\itme{\end{itemize}}
\def\enmb{\begin{enumerate}}
\def\enme{\end{enumerate}}
\def\eqnb{\begin{equation}}
\def\eqne{\end{equation}}

\def\PRL{Phys. Rev. Lett.}
\def\PRD{{Phys. Rev.} D}

\title{Cartan's Supersymmetry and \\Weak and Electromagnetic Interactions}  
\author{Sadataka Furui\\
 Graduate School of Teikyo University\\
2-17-12 Toyosatodai, Utsunomiya, 320-0003 Japan{\thanks
{\textit{E-mail address:} furui@umb.teikyo-u.ac.jp}}
}

\begin{document}
\maketitle
\begin{abstract}
We apply the Cartan's supersymmetric model to the weak interaction of hadrons.  
The electromagnetic currents are transformed by $G_{12},G_{123},G_{13},G_{132}$ and  the factor$(1-\gamma_5)$ is inserted between $l \bar\nu$ or $\bar l \nu$ when the photon is replaced by $W^\pm$, and between $l \bar l$ or $\nu\bar \nu$ when the photon is replaced by $Z$.

Electromagnetic currents in the Higgs boson $H^0$ decay into 2$\gamma$ and $D_s(0^+)$ decay into $D_s(0^-)\pi$ and $D_s(0^-)\gamma$ in which leptons are replaced by quarks are also studied.

A possibility that the boson near the $B\bar B$ theshold $\chi_b(3P, 10.53$ GeV) is the Higgs boson partner $h^0$ is discussed. 

We adopt Dirac lepton neutrinos and Majorana quark neutrinos, and construct a model that satisfy the $Z_3$ symmetry of the lepton sector and the quark sector, by adding two right-handed neutrinos whose left-handed partner cannot be detected by our electro-magnetic detectors.

%\keywords{Cartan's supersymmetry, Higgs boson $H^0$ and $h^0$, $D_s(0^+)$ boson decay}
\end{abstract}
\newpage
\section{Introduction}
 Cartan\cite{Cartan66} formulated the coupling of 4-dimensional spinors $A,B,C,D$ and 4-dimensional vectors $E,E'$ using the Clifford algebra, which is a generalization of quaternions and octonions. 
In this model there appears a triality symmetry, and one can imagine a presence of sectors of $E$ and $E'$ which cannot be detected by fermions in our detectors, in other words, fermions in our universe are transformed by $G_{12},G_{13},G_{123}$ and $G_{132}$ to vectors, but the vectors produced by these transformations cannot be detected by our electromagnetic probes. 

We applied the Cartan's supersymmetry to our physical system and applied to the decay of $\pi^0, \eta, \eta'$ to $\gamma\gamma$ \cite{SF12a,SF12b,SF12c,SF13a,SF13b}. The Pauli spinor was treated as a quaternion and the Dirac spinor was treated as an octonion. In the $\pi^0$ decay, the two final vector fields belong to the same group ($EE$) or ($E'E'$), and we call the diagram rescattering diagram.  In the decay of $\eta$, and $\eta'$, final vector fields belong to different groups ($EE'$), which we called twisted diagrams.  Qualitative difference of the $\eta$ decay or $\eta'$ decay, and $\pi^0$ decay can be explained by symmetry of Cartan's spinor.

The Clifford algebraic spinor of Cartan $\psi\in (\Vec C\bigotimes C\ell_{1,3})f$ is associated with the Dirac spinor
\begin{eqnarray}
\psi&=&\xi_1\Vec i+\xi_2\Vec j+\xi_3\Vec k+\xi_4
=\left(\begin{array}{cc}\xi_4+i\xi_3 &i \xi_1-\xi_2\\
                                i\xi_1+\xi_2&\xi_4-i\xi_3\end{array}\right)\nonumber\\
C\psi&=&-\xi_{234}\Vec i-\xi_{314}\Vec j-\xi_{124}\Vec k+\xi_{123}
=\left(\begin{array}{cc}\xi_{123}-i\xi_{124}&-i\xi_{234}+\xi_{314}\\
                                -i\xi_{234}-\xi_{314}&\xi_{123}+i\xi_{124}\end{array}\right)
\end{eqnarray}
and the spinor operator
\begin{eqnarray}
\phi&=&\xi_{14}\Vec i+\xi_{24}\Vec j+\xi_{34}\Vec k+\xi_0
=\left(\begin{array}{cc}\xi_0+i\xi_{34} &i \xi_{14}-\xi_{24}\\
                                i\xi_{14}+\xi_{24}&\xi_0-i\xi_{34}\end{array}\right)\nonumber\\
C\phi&=&-\xi_{23}\Vec i-\xi_{31}\Vec j-\xi_{12}\Vec k+\xi_{1234}
=\left(\begin{array}{cc}\xi_{1234}-i\xi_{12} &-i \xi_{23}+\xi_{31}\\
                                -i\xi_{23}-\xi_{31}&\xi_{1234}+i\xi_{12}\end{array}\right)
\end{eqnarray}

The trilinear form in these bases is
\begin{eqnarray}
{\mathcal F}&=&^t\phi CX\psi={^t\phi} \gamma_0x^\mu\gamma_\mu\psi\nonumber\\
&=&x^1(\xi_{12}\xi_{314}-\xi_{31}\xi_{124}-\xi_{14}\xi_{123}+\xi_{1234}\xi_1)\nonumber\\
&+&x^2(\xi_{23}\xi_{124}-\xi_{12}\xi_{234}-\xi_{24}\xi_{123}+\xi_{1234}\xi_2)\nonumber\\
&+&x^3(\xi_{31}\xi_{234}-\xi_{23}\xi_{314}-\xi_{34}\xi_{123}+\xi_{1234}\xi_3)\nonumber\\
&+&x^4(-\xi_{14}\xi_{234}-\xi_{24}\xi_{314}-\xi_{34}\xi_{124}+\xi_{1234}\xi_4)\nonumber\\
&+&x^{1'}(-\xi_{0}\xi_{234}+\xi_{23}\xi_{4}-\xi_{24}\xi_{3}+\xi_{34}\xi_2)\nonumber\\
&+&x^{2'}(-\xi_{0}\xi_{314}+\xi_{31}\xi_{4}-\xi_{34}\xi_{1}+\xi_{14}\xi_3)\nonumber\\
&+&x^{3'}(-\xi_{0}\xi_{124}+\xi_{12}\xi_{4}-\xi_{14}\xi_{2}+\xi_{24}\xi_1)\nonumber\\
&+&x^{4'}(\xi_{0}\xi_{123}-\xi_{23}\xi_{1}-\xi_{31}\xi_{2}-\xi_{12}\xi_3)
\end{eqnarray}

In the case of weak current, we replace the coupling $\gamma_0x^\mu\gamma_\mu$ to
$\gamma_0x^\mu\gamma_\mu(1-\gamma_5)$ and try to make the couplings between fermions and vector particles become unified in the form
\[
\sum_{i=1}^4 (x^i C\phi\, C\psi +x^{i'}C\phi\, \psi)
\]
by suitable choice of $1$ or $-\gamma_5$. 

Except the term $x^{4'}\xi_0\xi_{123}$, which is $x^{i'} \phi C\psi$ type, it is possible by the following choice
\begin{eqnarray}
{\mathcal G}&=&x^1(\xi_{12}\xi_{314}-\xi_{31}\xi_{124}+\langle\xi_{14}\gamma_5\rangle\xi_{123}-\xi_{1234}\langle\gamma_5\xi_1\rangle)\nonumber\\
&+&x^2(\xi_{23}\xi_{124}-\xi_{12}\xi_{234}+\langle\xi_{24}\gamma_5\rangle\xi_{123}-\xi_{1234}\langle\gamma_5\xi_2\rangle)\nonumber\\
&+&x^3(\xi_{31}\xi_{234}-\xi_{23}\xi_{314}+\langle\xi_{34}\gamma_5\rangle\xi_{123}-\xi_{1234}\langle\gamma_5\xi_3\rangle)\nonumber\\
&+&x^4(\langle\xi_{14}\gamma_5\rangle\xi_{234}+\langle\xi_{24}\gamma_5\rangle\xi_{314}+\langle\xi_{34}\gamma_5\rangle\xi_{124}-\xi_{1234}\langle\gamma_5\xi_4\rangle)\nonumber\\
&+&x^{1'}(\langle\xi_{0}\gamma_5\rangle\xi_{234}+\xi_{23}\xi_{4}-\xi_{24}\xi_{3}+\langle - \xi_{34}\gamma_5\rangle\xi_2)\nonumber\\
&+&x^{2'}(\langle\xi_{0}\gamma_5\rangle\xi_{314}+\xi_{31}\xi_{4}-\xi_{34}\xi_{1}+\langle- \xi_{14}\gamma_5\rangle\xi_3)\nonumber\\
&+&x^{3'}(\langle\xi_0 \gamma_5\rangle\xi_{124}+\xi_{12}\xi_{4}-\xi_{14}\xi_{2}+\langle- \xi_{24}\gamma_5\rangle\xi_1)\nonumber\\
&+&x^{4'}(\xi_0 \xi_{123}-\xi_{23}\xi_{1}-\xi_{31}\xi_{2}-\xi_{12}\xi_3).
\end{eqnarray}

When we define $\tilde\xi_{i}=\xi_{jkl}$, $1\leq j,k,l\leq 3, j,k,l\ne i$, $\tilde\xi_{ij}=\xi_{kl}$, $1\leq k,l\leq 4, k,l\ne i,j$, $\tilde\xi_{ijk}=\xi_{\l}$,
$1\leq l\leq 4$ and $l \ne i,j,k$, $\tilde \xi_0=\xi_{1234}$ and $\tilde \xi_{123}=\xi_4$, the couplings that can be detected becomes 
\begin{eqnarray}
{\mathcal G}&=&\langle {^t\phi} CX(1-\gamma_5)\psi\rangle=\langle{^t\phi} \gamma_0x^\mu\gamma_\mu(1-\gamma_5)\psi\rangle\nonumber\\
&=&\sum_{i=1}^3 x^i(\xi_{i[i+1]_3}\xi_{[i+2]_3 i 4} -\xi_{[i+2]_3 i}\xi_{i[i+1]_3 4} -\tilde\xi_{i4}\xi_{123} +\xi_{1234}\tilde\xi_{i})\nonumber\\
&&+x^4(-\tilde\xi_{14}\xi_{234}-\tilde\xi_{24}\xi_{314}-\tilde\xi_{34}\xi_{124} +\xi_{1234}\tilde\xi_4) \nonumber\\
&+&\sum_{i=1}^3 x^{i'}(-\tilde \xi_0\xi_{[i+1]_3 [i+2]_3 4} +\xi_{[i+1]_3 [i+2]_3}\xi_4-\xi_{[i+1]_3 4}\xi_{[i+2]_3}+\tilde\xi_{[i+2]_34}\xi_{[i+1]_3})\nonumber\\
&&+x^{4'}(\xi_0\xi_{123}-\xi_{23}\xi_1-\xi_{31}\xi_2-\xi_{12}\xi_3).
\end{eqnarray}
Here the notation $[i+k]_3$ stands for $Mod[i+k, 3]$.  

If one multiplies $-\gamma_5$ to the exceptional term $x^{4'}\xi_0\xi_{123}$, the term becomes $x^{4'}C\phi C\psi$ type, and  since  there is no difference between $x^{4'}$ and $x^4$ in the electromagnetic interaction, the weak interaction can be characterized  as
\[
{^t \phi} CX C\psi \, +{^t \phi} CX \psi.
\]
The states $\psi$ and $C\psi$ makes a complete set of the initial state, and final states of our weak interactions is $\phi$.

In sect.2, we present Lagrangian of the weak interaction and define the Higgs field.
Electromagnetic decays of Higgs bosons are studied in sect.3.  Electromagnetic decays and weak decays of $B(0^+)$ bosons and $D_s(0^+)$ bosons are studied in sect.4, and discussion and conclusion are given in sect.5.

\section{Weak interaction of leptons and hadrons}
In the case of the vector particle $x^i$ and $x^{i'}$ are $W^+$, we choose $C\phi=\nu$, $\phi=\bar\nu$  (neutrino and antineutrino), and $C\psi=\bar \l$, $\psi=\l$ (antilepton and lepton) or $C\psi=\bar q$, $\psi=q$ (antiquark and quark). 

In the case of the vector particle $x^i$ and $x^{i'}$ are $Z$, we choose the spinor $^t(\psi, C\psi)$ and $^t(\phi, C\phi)$ the two quarks in one triaity sector, or lepton antilepton pairs and neutrino  antineutrino pairs. 

The spinor of a neutrino is defined as 
\[
\Psi_p=\left(\begin{array}{c}\eta_p\\
                                          \chi_p\end{array}\right)=
\left(\begin{array}{c}i\sigma^2\chi_p^{\dagger T}\\
                                   \chi_p\end{array}\right),
\]
where we define
\[
\sigma^1=\left(\begin{array}{cc}0&1\\
                                          1&0\end{array}\right)\quad 
\sigma^2=\left(\begin{array}{cc}0&-i\\
                                          i&0\end{array}\right)\quad 
\sigma^3=\left(\begin{array}{cc}1&0\\
                                          0&-1\end{array}\right) .
\]
and 
\[
\sigma^\mu=(\Vec 1,\Vec \sigma) \qquad
\bar\sigma^\mu=(\Vec 1,-\Vec \sigma).
\]
and 
\begin{eqnarray}
&&(E\Vec 1-\Vec \sigma\cdot \Vec p)\eta_p =m\chi_p\nonumber\\
&&(E\Vec 1+\Vec \sigma\cdot \Vec p)\chi_p=m\eta_p
\end{eqnarray}

The Lagrangian of a Majorana neutrino is given as\cite{Labelle10}
\[
\mathcal L_M=\chi_p^\dagger \bar\sigma^\mu i\partial_\mu \chi_p-\frac{m}{2}(\chi_p\cdot\chi_p+\bar\chi_p\cdot\bar\chi_p),
\]
and we adopt the lepton neutrinos are Dirac neutrinos and quark neutrinos are Majorana neutrinos\cite{SV80}.

The leptons, quarks and neutrinos have triality sectors
\[
(b,t|\tau, \nu_\tau), \quad (s,c|\mu, \nu_\mu),\quad (u,d|e, \nu_e).
\]

The neutrinos $\nu_e, \nu_\mu$ and $\nu_\tau$ are described by a unitary transformation $U$ on $\nu_1,\nu_2,\nu_3$ as\cite{PDG14}
\[
\left(\begin{array}{c} \nu_e\\
                               \nu_\mu\\
                               \nu_\tau\end{array}\right)=
\left(\begin{array}{ccc}c_{12}c_{13}& s_{12}c_{13}& s_{13}e^{-i\delta}\\
          -s_{12}c_{23}-c_{12}s_{23}s_{13}e^{i\delta}&c_{12}c_{23}-s_{12}s_{23}s_{13} e^{i\delta}&s_{23}c_{13}\\
       -s_{12}s_{23}-c_{12}c_{23}s_{13}e^{i\delta}&-c_{12}s_{23}-s_{12}c_{23}s_{13}e^{i\delta}&c_{23}c_{13}\end{array}
\right)\left(\begin{array}{c} \nu_1\\
                                       e^{i\frac{\alpha_{21}}{2}} \nu_2\\
                                        e^{i\frac{\alpha_{31}}{2}} \nu_3\end{array}\right )
\] 
where $c_{ij}=\cos\theta_{ij}$ and $s_{ij}=\sin\theta_{ij}$, and  $\theta=[0,\pi/2]$.
The angle $\delta=[0,2\pi]$ is the Dirac CP violating phase, and $\alpha_{21}$ and $\alpha_{31}$ are Majorana CP violating phases.

In the minimal supersymmetric standard model(MSSM), there is a Higgs doublet which is even under R-parity and quarks and leptons which are odd under R-parity. When a system has baryon number $B$, lepton number $L$ and spin $s$, the R-parity becomes
\[
R=(-1)^{3(B-L)+2s}
\] 
which can be regarded as a consequence of flavor symmetry $\Delta(27)$\cite{MPV13}. 

The $\Delta(27)$ symmetry prefers degeneracy of $\nu_e,\nu_\mu$ and $\nu_\tau$, however the Higgs model prefers two degenerate light and one heavy neutrino\cite{SF12c}. Violation of $\Delta(27)$ symmetry is large also in the quark neutrino sector. 

Since the mass of the top quark 174 GeV is heavier than the Higgs boson mass 125 GeV, mass terms of neutrinos in systems containing other heavior Higgs doublets were studied\cite{ZHR09n,ZHR10n}, and the stability of the vacuum which contains arbitrary number of CP violating Higgs doublets was studied \cite{ZHR09,ZHR10}. 
The heavy top quark mass does not necessarily require heavior Higgs doublets, since we do not know whether the mass of the top quark is fixed by the Yukawa coupling.
  
In the standard model, leptons are defined by left-chiral field 
\[
\Psi_L=\left(\begin{array}{c}\Psi_{l_L}\\
                                     \Psi_{\nu_{l L}}\end{array}\right) 
\]
whose covariant derivative is
\[
D_\mu\Psi_L=(\partial_\mu+i g \frac{\tau^i}{2}{W^i}_\mu-i\frac{g'}{2}B_\mu)\Psi_L
\]
and right-chiral field $\Psi_{l R}$, whose covariant derivative is
\[
D_\mu\Psi_{l R}=(\partial_\mu-i g'B_\mu)\Psi_{l R}
\]
The weak bosons $W$ are weighted by isospin operator $\tau^i/2$, and the gluons $G$ are weighted by the color operator $\lambda^a/2$.

The covariant derivative of left-handed quark field is
\[
D_\mu\Psi_{Q_L}=(\partial_\mu+i g_s\frac{\lambda^a}{2}{G^a}_\mu+ig\frac{\tau^i}{2}{W^i}_\mu+i\frac{g'}{6}B_\mu)\Psi_{Q_L}.
\]
The covariant derivative of $u_R$ and $d_R$ quarks are
\begin{eqnarray}
D_\mu\Psi_{u_R}&=&(\partial_\mu+i g_s\frac{\lambda^a}{2}G^a_\mu+i\frac{2g'}{3}B_\mu)\Psi_{u_R}\nonumber\\
D_\mu\Psi_{d_R}&=&(\partial_\mu+i g_s\frac{\lambda^a}{2}G^a_\mu-i\frac{g'}{3}B_\mu)\Psi_{d_R}
\end{eqnarray}

The Lagrangian of the $U(1)_Y, SU(2)_L$ and $SU(3)_C$ gauge fields is
\[
{\mathcal L}_{GB}=-\frac{1}{4}F^{\mu\nu}F_{\mu\nu}-\frac{1}{2}Tr(G_{\mu\nu}G^{\mu\nu})
\]
where
\begin{eqnarray}
F_{\mu\nu}&=&\partial_\mu B_\nu-\partial_\nu B_\mu\nonumber\\
W_{\mu\nu}&=&\partial_\mu W_\nu-\partial_\nu W_\mu-ig[W_\mu, W_\nu]\nonumber\\
G_{\mu\nu}&=&\partial_\mu G_\nu-\partial_\nu G_\mu-ig[G_\mu, G_\nu]
\end{eqnarray}

The Higgs field $H$ belongs to $SU(2)_L$ doublet and can be expressed as
\[
H_u=\left(\begin{array}{c} H_u^+\\
                                H_u^0\end{array}\right), \quad
H_d=\left(\begin{array}{c} H_d^0\\
                                H_d^-\end{array}\right), 
\]
The gauge-covariant derivative $D_\mu H$ is 
\[
D_\mu H=(\partial_\mu+ig\frac{\tau^i}{2}W^i_\mu+i\frac{g'}{2}B_\mu)H
\]

In our model, the Higgs potential is written as
\[
V(H)=-\mu^2 H^\dagger H+\lambda(H^\dagger H)^2
\]
and the vacuum expectation value of $H$ is chosen to be
\[
\langle H_u\rangle_{min}=\left(\begin{array}{c} 0\\
                                                  \nu_u\end{array}\right), \quad
\langle H_d\rangle_{min}=\left(\begin{array}{c} \nu_d\\
                                                  0\end{array}\right), 
\]
where $\displaystyle\nu_u=\frac{\mu_u}{\sqrt{2\lambda}}$, $\displaystyle\nu_d=\frac{\mu_d}{\sqrt{2\lambda}}$, $\mu_u^2+\mu_d^2=2\mu^2$ and define 
\[
{H_0}_u'=H_0-\nu_u \quad\rm{ and} \quad {H_0}_d'=H_0-\nu_d. 
\]
The covariant derivative of $H_u$ becomes
\[
D_\mu H_u=D_\mu\left(\begin{array}{c}H_u^+\\
                                                 {H_0}_u'\end{array}\right)
+D_\mu\left(\begin{array}{c} 0\\
                                  \nu_u\end{array}\right)
\]
where
\[
D_\mu\left(\begin{array}{c} 0\\
                                  \nu_u\end{array}\right)=
\frac{i\nu_u}{2}\left(\begin{array}{c} gW_{1\mu}-ig W_{2\mu}\\
                                             -gW_{3\mu}+g' B_\mu\end{array}\right).
\]

Similarly
\[
D_\mu H_d=D_\mu\left(\begin{array}{c}{H_0}_d'\\
                                                 H_d\end{array}\right)
+D_\mu\left(\begin{array}{c} \nu_d\\
                                  0\end{array}\right)
\]
where
\[
D_\mu\left(\begin{array}{c} \nu_d\\
                                   0\end{array}\right)=
\frac{i\nu_d}{2}\left(\begin{array}{c} gW_{3\mu}\\
                                      gW_{1\mu}+ig W_{2\mu}-g' B_\mu\end{array}\right).
\]

The masses of the gauge bosons are
\begin{eqnarray}
{\mathcal L}_{MGB}&=&\frac{\nu^2}{4}g^2(W_1^2+W_2^2)+\frac{\nu^2}{4}(gW_3-g' B)^2\nonumber\\
&=&m_W^2 W^{+\mu}{W^-}_\mu+\frac{1}{2}m_Z^2 Z_\mu Z^\mu,
\end{eqnarray}
where $\nu^2=\nu_u^2+\nu_d^2$.

The possible masses of the Higgs particles are
\[
m_{H^\pm}^2=m_W^2+m_{A^0}^2
\]
for charged massive states, and
\begin{eqnarray}
m^2_{h_0}&=&\frac{m_{A_0}^2+m_Z^2}{2}-\frac{1}{2}\sqrt{(m_{A_0}^2+m_Z^2)^2-4m_{A_0}^2m_Z^2\cos^2 2\beta}\nonumber\\
m^2_{H_0}&=&\frac{m_{A_0}^2+m_Z^2}{2}+\frac{1}{2}\sqrt{(m_{A_0}^2+m_Z^2)^2-4m_{A_0}^2m_Z^2\cos^2 2\beta},
\end{eqnarray}
where $\tan\beta=\nu_u/\nu_d$, for neutral massive states.

When $\cos 2\beta=0$, $m^2_{h_0}=0$, $m^2_{H_0}=m_{A_0}^2+m_Z^2$, and $m_Z=91.2$GeV, $m_{H_0}=125$GeV\cite{ATLAS12a} gives 
\[
m_{A_0}=85.5 \rm{GeV}.
\]
and $m_W=80.4$ GeV yields $m_H^\pm=117$ GeV.

There is a search of $H^+$ through the decay of the top quark to $H^+b$ and through the analysis of $H^+\to\tau\nu_\tau$, $m_{H^+}=120$GeV was obtained\cite{CMS12}. In this analysis, the branching fraction $B(H^+\to\tau\nu_\tau$ could not be fixed and it was assumed to be 1. We expect that the instability of  the $H^+$ state is due to supersymmetry breaking potential. The requirement that ${m_H^\pm}=\sqrt{m_W^2+m_{A_0}^2}=120$ GeV gives $m_{A0}=95.7$ GeV and $m_H^0$ becomes 125 GeV, with 
\[
m_{A0}=95.7 \rm{GeV \quad and} \quad \cos 2\beta=\pm 0.1687.
\]

The fixed $\cos 2\beta$ gives $m_{h0}= 11.2$ GeV, and near this energy region there are $\chi_{b0}(1P, J^{PC}=0^{++}, 9.86$ GeV), $\chi_{b1}(1P, J^{PC}=1^{++}, 9.89$ GeV) and  $\chi_{b2}(2P, J^{PC}=2^{++}, 10.23$ GeV) which are expected to be made of $b\bar b$ and 
$\chi_b(3P , 10.53$ GeV).  The scalar boson $\chi_b(3P)$ decays radiatively to $\Upsilon$(1S) and $\Upsilon$(2S), and its $C=+$ but its $J^P$ is not well known\cite{ATLAS12b,Abazov12}.

 The mass of $\chi_b(3P)$ is slightly below the $B\bar B$ threshold and there remains a possibility that the SUSY-breaking potential\cite{Labelle10},
\begin{eqnarray}
V_{SSB}&=&b (H_u^+  H_u^0) i\tau_2 \left(\begin{array}{c} H_d^0\\
                                                                                 H_d^-\end{array}\right)
            +b^*(H_d^{0\dagger} H_d^{-\dagger})(- i\tau_2)\left(\begin{array}{c} H_u^{+\dagger}\\
                                                              H_u^{0\dagger}\end{array}\right)\nonumber\\
         &=&b(H_u^+ H_d^- -H_u^0 H_d^0) +h.c.
\end{eqnarray}
where $b>0$, makes $m_H^\pm$ unstable, and the $m_{h0}$ appears as that of the $\chi_b(3P)$.  
Detailed study of the structure of $\chi_b(3P)$ is necessary to clarify the Higgs meson physics. 
 
\section{$H(0^+)\to \ell \bar \ell \ell \bar\ell \to 2\gamma$}
\begin{figure}[htb]
\begin{minipage}[b]{0.47\linewidth}
\begin{center}
\includegraphics[width=5cm,angle=0,clip]{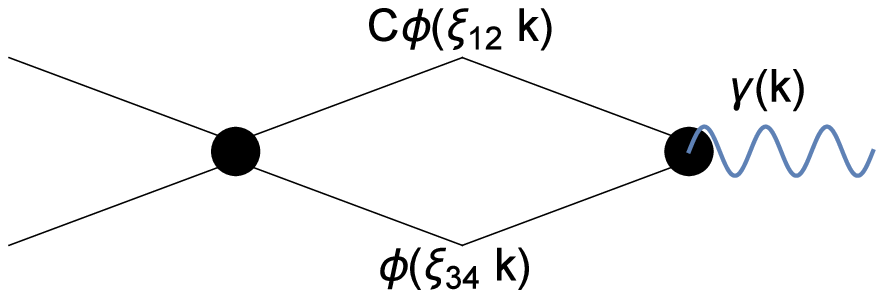}
%\label{g1a}
\end{center}
\end{minipage}
\hfill
\begin{minipage}[b]{0.47\linewidth}
\begin{center}
\includegraphics[width=5cm,angle=0,clip]{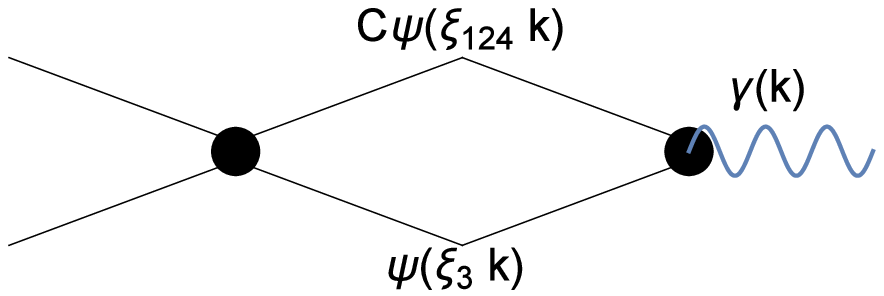}
%\label{g1i}
\end{center}
\end{minipage}
\caption{Typical diagrams of  lepton-antilepton $\phi-{\mathcal C}\phi$ or $\psi-{\mathcal C}\psi$ decay into a $\gamma$. }
\label{gl}
\end{figure}

As a model of $H(0^+)$, we use spinor fields in Clifford algebra\cite{Lo01, He86}, and study its decay into 2$\gamma$.
 In the lepton-antilepton annihilation and quark-antiquark annihilation to $\gamma$, the helicity of the lepton and the antilepton are assumed to be parallel. 
 Typical diagrams of $\phi-{\mathcal C}\phi$ or $\psi-{\mathcal C}\psi$ decays into a $\gamma$ in the standard model are shown in Figure \ref{gl}.

 In the octonion bases, in addition to the lepton-antilepton pair $\phi-{\mathcal C}\phi$ or $\psi-{\mathcal C}\psi$, the pair $\phi-\psi$ or ${\mathcal C}\phi-{\mathcal C}\psi$ decays into a $\gamma$. 

\begin{figure}[h]
\begin{minipage}[b]{0.47\linewidth}
\begin{center}
\includegraphics[width=5cm,angle=0,clip]{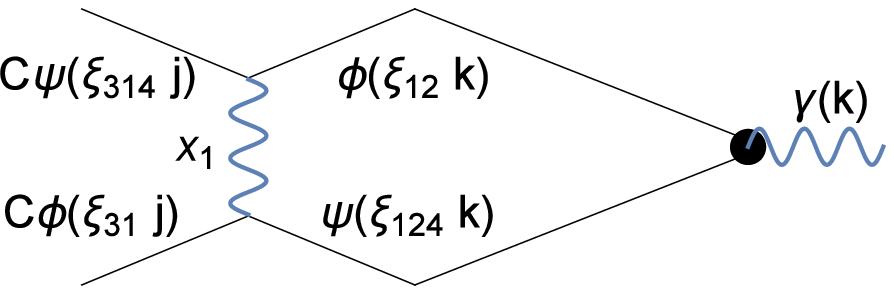}
%\label{g1a}
\end{center}
\end{minipage}
\hfill
\begin{minipage}[b]{0.47\linewidth}
\begin{center}
\includegraphics[width=5cm,angle=0,clip]{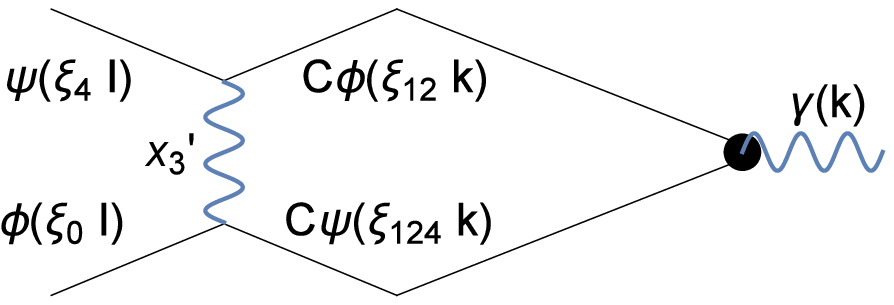}
\label{gla}
\end{center}
\end{minipage}
\caption{Typical diagrams of lepton-antilepton $\phi-\psi$ or ${\mathcal C}\phi-{\mathcal C}\psi$ decay into a $\gamma$. }
\end{figure}

%The Clifford algebraic spinor $\psi\in (\Vec C\bigotimes C\ell_{1,3})f$ are associated with the spinor $\Phi$ or the spinor $\Psi$, defined in the introduction.

A scalar boson $\Psi(0^+)$ or $\Phi(0^+)$ decays into $\gamma(\ell \bar\ell)\gamma(\ell \bar \ell)$, where $\ell$ stands here for $e$ or $\mu$. There are 8 diagrams each\cite{SF14}

\begin{figure}[htb]
\begin{minipage}[b]{0.47\linewidth}
\begin{center}
\includegraphics[width=5cm,angle=0,clip]{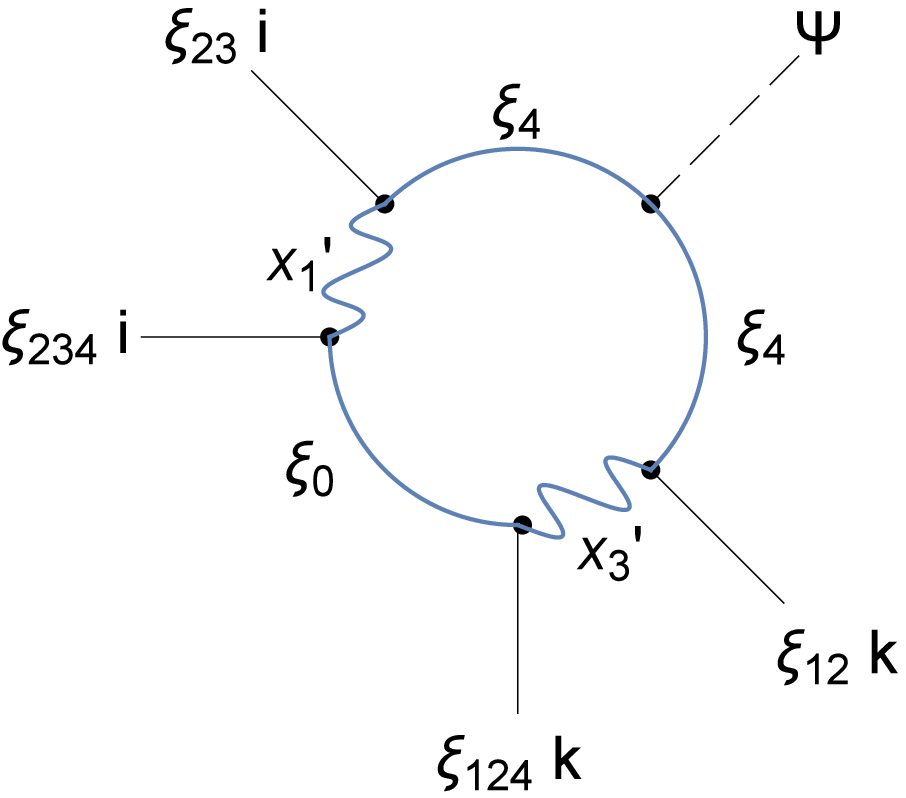}
%\label{gb1e}
\end{center}
\end{minipage}
\hfill
\begin{minipage}[b]{0.47\linewidth}
\begin{center}
\includegraphics[width=5cm,angle=0,clip]{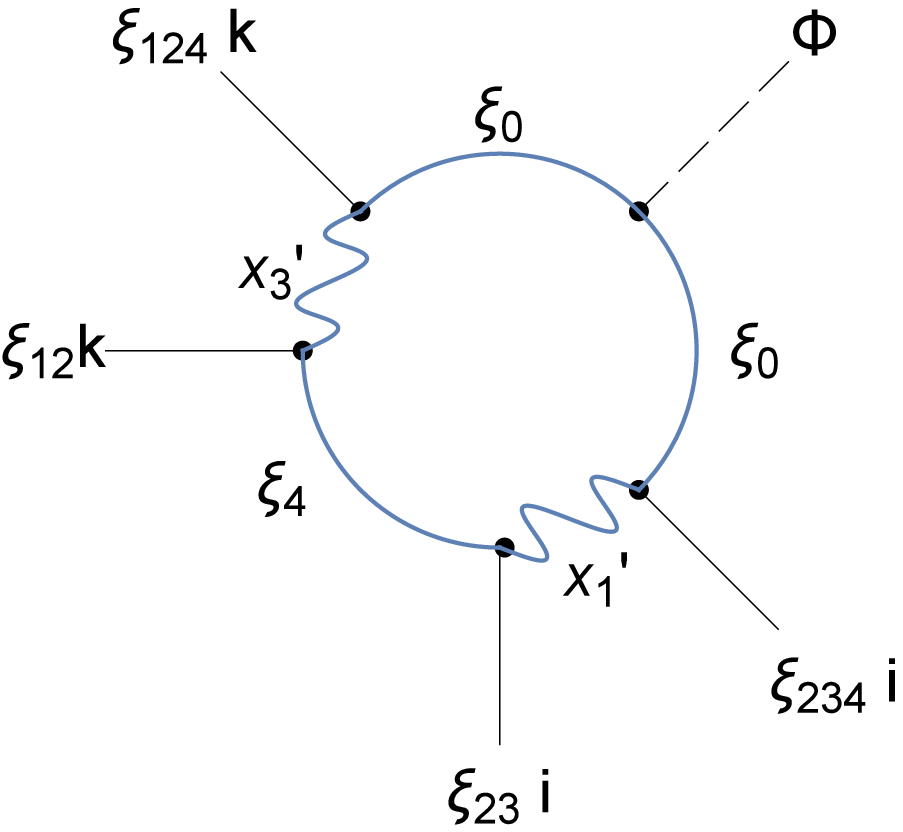}
%\label{gb1f}
\end{center}
\end{minipage}
\caption{Typical diagrams of  $H(0^+)$ decay into $\ell\bar\ell \ell\bar\ell$ which reduce to 2$\gamma$s. }
\label{gb1}
\end{figure}

In the Figure \ref{gb1}, $\xi_*$s and $\xi'_*$s stand for leptons or antileptons that decay into $\gamma$s and $\xi_i$s stand for photons or gauge bosons.
We studied decay modes of $H_0$ in \cite{SF14}.

%\newpage
\section{$B(0^+)\to B(0^-)\pi$ and $B(0^+)\to D_s(0^+)\mu^+\nu_\mu$} 
Experimentally, presence of charmed strange meson $D_s(0^+)$ was exciting and the presence of bosonized strange meson $B_s(0^+)$ and hadronic decays of a $B(0^+)$ meson was studied in lattice simulation\cite{Green03, McNeil04}. 
Experimentally the $B^+(0^+)$ decay into $B(0^-)\pi^+$ is not observed. However, $D_s^+(0^+)$  decay into $D_s^+(0^-)\pi^0$ and $D_s^+(0^-)\gamma$ are observed\cite{PDG14} 
\begin{eqnarray}
&&D_s^+(0^-) (1968{\rm MeV}) \pi^0\quad 5.8\pm 0.7\% \nonumber\\
&&D_s^+(0^-) (1968{\rm MeV}) \gamma\quad 94.2\pm 0.7\%. \nonumber
\end{eqnarray}
 Quarks and anti quarks are expressed by $\phi, {\mathcal C}\phi, \psi$ and ${\mathcal C}\psi$.
 In the analysis of Higgs boson decay into 2$\gamma$s,  $\phi, \psi$ and $ {\mathcal C}\phi, {\mathcal C}\psi$ pair creation/annihilation was consistent with $\gamma$ creation/annihilation\cite{SF14}. 

\begin{figure}[htb]
\begin{minipage}[b]{0.47\linewidth}
\begin{center}
\includegraphics[width=5cm,angle=0,clip]{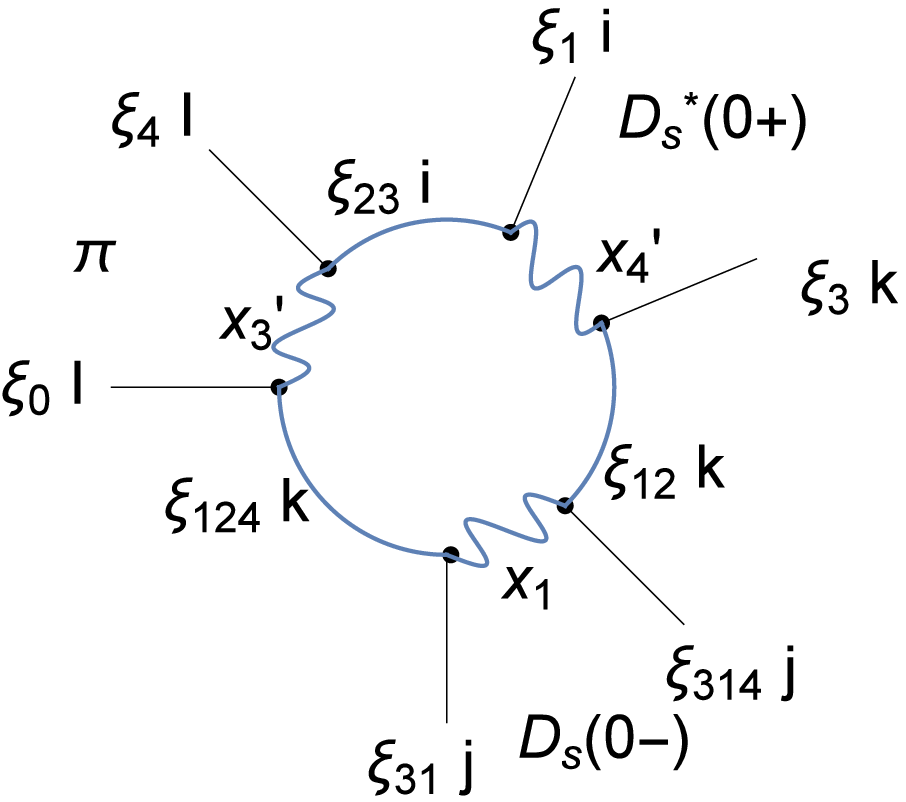} %{D_DPi_jk112p.eps}
%\label{DDPi7}
\end{center}
\end{minipage}
\hfill
\begin{minipage}[b]{0.47\linewidth}
\begin{center}
\includegraphics[width=5cm,angle=0,clip]{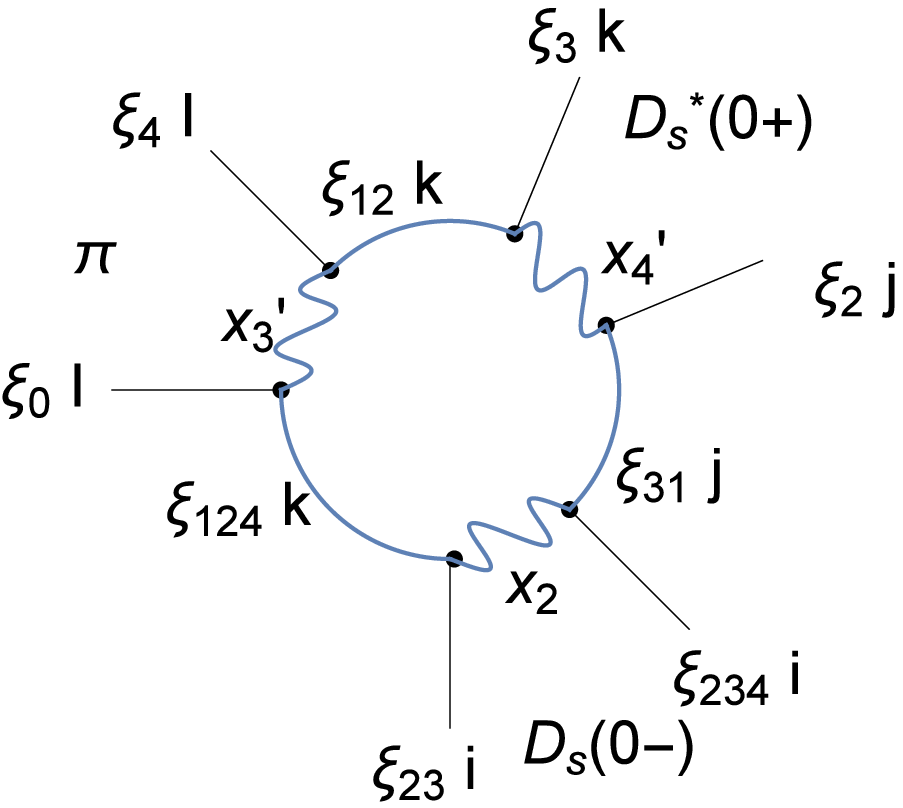}  %{D_DPi_ik223p.eps}  
%\label{DDpi8}
\end{center}
\end{minipage}
\caption{Typical diagrams of $D_s^*(0^+)\to D(0^-)+\pi(q (\xi_4)\bar q(\xi_0) )$.}
 \label{DDpi1}
\end{figure}

\begin{figure}
\begin{minipage}[b]{0.47\linewidth}
\begin{center}
\includegraphics[width=5cm,angle=0,clip]{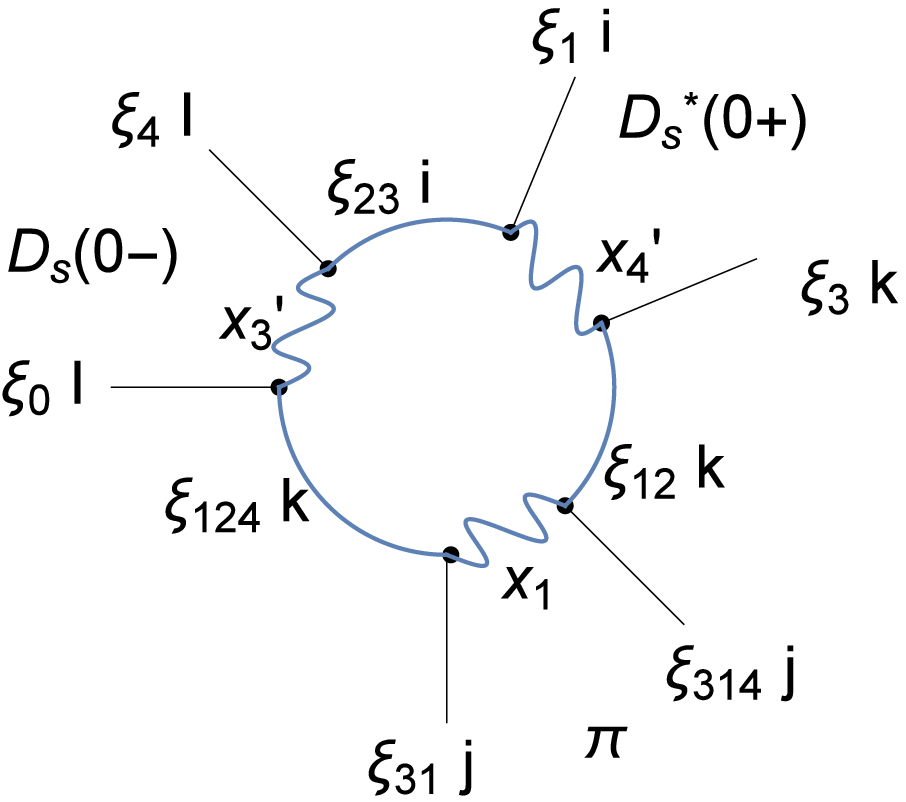} %{D_DPi_ik221p.eps}
%\label{DDPi9}
\end{center}
\end{minipage}
\hfill
\begin{minipage}[b]{0.47\linewidth}
\begin{center}
\includegraphics[width=5cm,angle=0,clip]{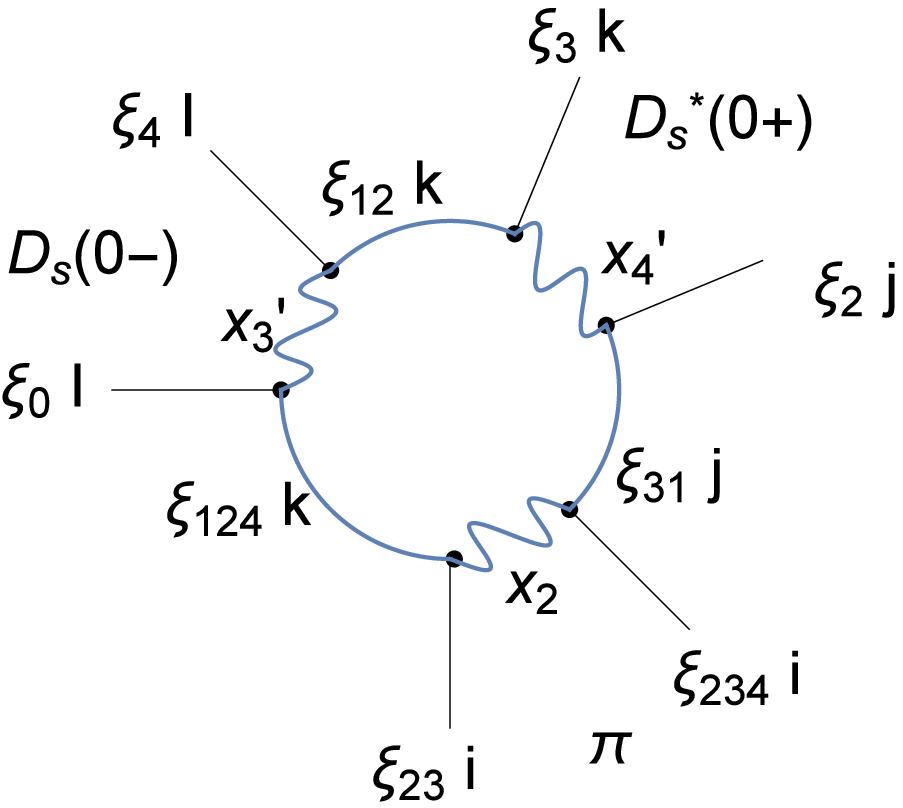} % {D_DPi_jk113p.eps}
%\label{DDpi10}
\end{center}
\end{minipage}
\caption{Typical diagrams of $D_s^*(0^+)\to D(0^-)+\pi(q(\xi_{**}) \bar q(\xi_{***}) )$.}
 \label{DDpi2}
\end{figure}

The photons in the $D_s(0^+)\to D_s(0^-)\gamma$ appear from the interactions 
\begin{eqnarray}
&& {\mathcal C}\phi^*\gamma_0\gamma_\mu\phi, \qquad
 {\mathcal C}\psi^*\gamma_0\gamma_\mu\psi\nonumber\\
&& \phi^*\gamma_0\gamma_\mu\psi+h.c.\quad
{\mathcal C}\phi^*\gamma_0\gamma_\mu{\mathcal C}\psi+h.c.\nonumber
\end{eqnarray}
The pion is a quark-antiquark system coupled to angular momentum 0, and in the $D_s^*(0^+)\to D_s(0^-)\pi$, they appear from the divergence of an axial current
\[
{\mathcal C}\phi^*\gamma_0\gamma_5 \phi.
\]
 In order that quarks and antiquarks in $D_s^*(0^+)$ mesons are expressed by  $\xi_*\Vec i, \xi_*\Vec j, $ or $\xi_* \Vec k$, the polarization of quark-antiquarks in pions or $D(0^-)$ becomes $q(\Vec I)\bar q(\Vec I)$, as shown in Figures \ref{DDpi1} and \ref{DDpi2}.  We take interactions between quarks in $D(0^+)$ to be Coulomb type, in order that the relative p-wave becomes well determined, and distinguish them from that in $D(0^-)$.

\begin{figure}[htb]
\begin{center}
\includegraphics[width=8cm,angle=0,clip]{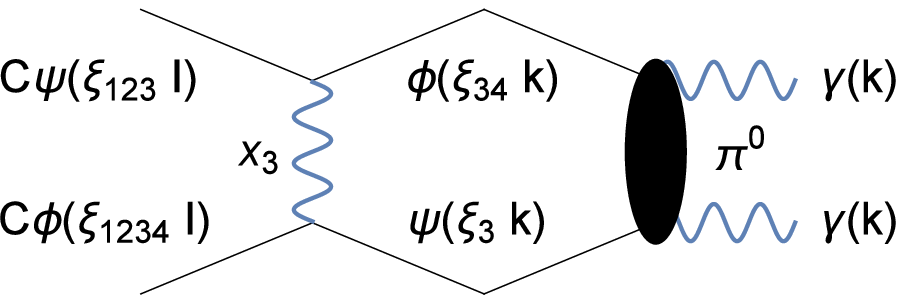}
\label{g1i}
\end{center}
\caption{ A quark-antiquark $\phi-\psi$ or ${\mathcal C}\phi-{\mathcal C}\psi$ decay into $2\gamma$ via $\pi^0$. }
\end{figure}

 When the helicities of the quark-antiquark are not parallel, decay into 2$\gamma$ via  $a_2(2^+)$ may occur.

\begin{figure}[htb]
\begin{center}
\includegraphics[width=8cm,angle=0,clip]{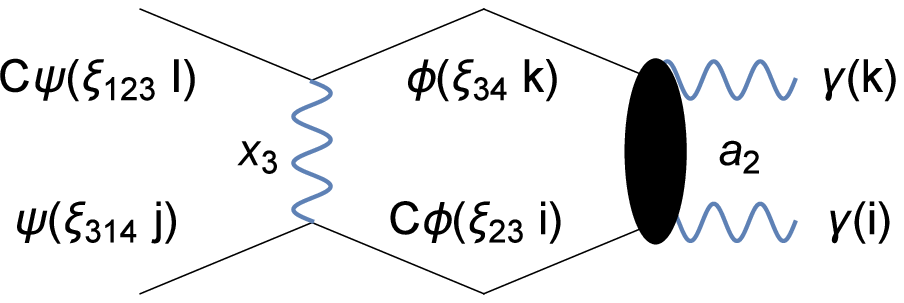}
\label{g2i}
\caption{ A quark-antiquark $\phi-{\mathcal C}\phi$ or $\psi-{\mathcal C}\psi$ decay into $2\gamma$ via $a_2$.} 
\end{center}
\end{figure}

\vskip 0.2 true cm

In the Figures \ref{DDpi1} and \ref{DDpi2}, $x_i$s and $x'_i$s stand for gluons.  The lines $\xi_{12}\Vec k$ and $\xi_{31}\Vec j$ in the Figure \ref{DDpi1}, and the lines $\xi_{23}\Vec i$ and $\xi_{12}\Vec k$ in the Figure \ref{DDpi2} are s-quarks.  

 Corresponding to $D_s(0^-)\pi^0\to D_s(0^-) 2\gamma$ decay, in $D_s(0^-)\gamma$ decay, $q(\phi)\bar q (\psi)\to \gamma$ and $q({\mathcal C}\phi)\bar q ({\mathcal C}\psi)\to \gamma$ occur, as shown in Figure \ref{gl}, i.e. the number of quarks per one $\gamma$ that contribute to $D_s(0^-)\gamma$ decay is 4 times larger than that contribute to $D_s(0^-)\pi$ decay.
The decay width of $D_s(0^-)\gamma$ would become 16 times larger than that of $D_s(0^-)\pi$.

Absence of  $B_s(0^+)$ decay into $B_s(0^-)\pi^+$ decay is expected to be due to the presence of triality sector $(s,c|\mu, \nu_\mu)$, which makes the weak decay of $B(0^+)\to D_s^*(0^+)\mu^+\nu_\mu$ stronger than the strong decay of $B(0^+)\to B(0^-)\pi^+$. A $D_s^*(0^+)$ decays to $D_s(0^-)+\pi$, and a $D_s(0^-)$ decays to $K$+anything by about 57 \% via strong interactions.

%\newpage
\section{Discussion and conclusion}
We showed that Cartan's supersymmetry can be applied to weak interactions of leptons and hadrons. Consistency with the electromagnetic interaction was also confirmed.
 
The model of Higgs boson predicts presence of two neutral scalar bosons of masses $m_{H^0}$ and $m_{h0}$ and charged  scalar boson of mass $m_{H^\pm}$. An adjustment of $m_{H^0}=m_{H^\pm}=125$ GeV predicts $m_{h0}\simeq 11$ GeV. There are possibility that $H^\pm$ is unstable and not observed, but the boson $\chi_b(3P, 10.53$ GeV) \cite{ATLAS12b,Abazov12} near the $B\bar B$ theshold is the Higgs boson partner $h^0$. 

Detailed study of $\chi_{b}(3P, 10.53$ GeV) decay may be helpful for clarifying whether the  $\chi_{b}(3P)$ can be understood as an $h_0$.

The world of matters transformed by $G_{23}$ can be understood through our detectors, and the world of matters transformed by $G_{12},G_{13},G_{123}$ and $G_{132}$ would be understood indirectly through behaviors of neutrinos, $H^0$ and $h^0$.

Application of Clifford algebra-valued differential form to the field strength effective to the Dirac spinor of  spin 1/2 and to the Rarita-Schwinger spinor of spin 3/2 was studied in \cite{Mielke01}. In the case of Dirac spinors, our model contains the triality symmetry, which can be interpreted as the color degrees of freedom ($r,g,b$). The flavor neutrinos are 
\[
\nu_e,  [\nu_u,\nu_d]_{r,g,b},\quad \nu_\mu, [\nu_c,\nu_s]_{r,g,b},\quad \nu_\tau, [\nu_t,\nu_b]_{r,g,b}
\]
all left-handed, and there are seven right-handed neutrinos. . 

Branco et al\cite{BRS13} proposed $Z_3$ symmetry among leptons and quarks, and introduced heavy right-handed Dirac and Majorana neutrinos for deriving light neutrinos in quark sectors and lepton sectors via the so-called seesaw mechanism\cite{SV80}.
 Aranda et al.\cite{ABMPV14} discussed that in order to create neutrino masses in a system with the $Z_3$ symmetry, neutrinos should be Dirac neutrinos instead of  Majorana neutrinos.

We do not adopt the $Z_3$ symmetry as \cite{BRS13}, but corresponding to the color degrees of freedom of quarks, we assume presence of six degrees of freedom in lepton sector,
\[
|e,\nu_e)^*,\quad  |\mu,\nu_\mu)^*, \quad |\tau,\nu_\tau)^*, \quad
|e,\nu_e)^{**},\quad |\mu,\nu_\mu)^{**}, \quad |\tau,\nu_\tau)^{**}
\]
which cannot be detected by our electro-magnetic detectors. Charged Higgs boson $H^{\pm}$ can connect to these states but the final states cannot be detected, as the experiment suggests\cite{CMS12}. We think the $\mu\to e\gamma$ cannot be detected by our electromagnetic detectors in contrast to \cite{ABMPV14}.

The number of massive neutrinos becomes $21+6=27$, and the number of right-handed neutrinos becomes $7+2=9$ in this model. It is possible to construct a model satisfying $Z_3$ symmetry using Dirac lepton neutrinos and Majorana quark neutrinos. 

\vskip 0.5 true cm
\leftline{\bf{Acknowledgements}}
The author thanks Dr. Fabian Cruz for sending the information of the ref.\cite{CMS12}, and the referees for attracting author's attention to the refs. \cite{ZHR09n,ZHR10n,ZHR09,ZHR10,Mielke01,ABMPV14}.
 
%\end{acknowledgements}
\newpage
% BibTeX users please use
%\bibliographystyle{spbasic}


\end{document}